\documentclass[twocolumn,twoside,slac_two]{revtex4}
\usepackage{graphicx}
\usepackage{fancyhdr}
\begin{document}

\title{Fractal dimensions of the galaxy distribution varying by steps?}


\author{M. N. C\'el\'erier}
\affiliation{Laboratoire Univers et TH\'eories (LUTH) Observatoire 
de Paris-Meudon, 5 place Jules Janssen, 92195 Meudon, FRANCE}
\author{R. Thieberger}
\affiliation{Physics Department, Ben Gurion University, Beer Sheva 84105, ISRAEL}

\begin{abstract}

The structure of the large scale distribution of the galaxies have been widely studied since the publication of the first catalogs. Since large redshift samples are available, their analyses seem to show fractal correlations up to the observational limits. The value of the fractal dimension(s) calculated by different authors have become the object of a large debate, as have been the value of the expected transition from fractality to a possible large scale homogeneity. Moreover, some authors have proposed that different scaling regimes might be discerned at different lenght scales. To go further on into this issue, we have applied the correlation integral method to the wider sample currently available. We therefore obtain a fractal dimension of the galaxy distribution which seems to vary by steps whose width might be related to the organization hierarchy observed for the galaxies. This result could explain some of the previous results obtained by other authors from the analyses of less complete catalogs and maybe reconcile their apparent discrepancy. However, the method applied here needs to be further checked, since it produces odd fluctuations at each transition scale, which need to be thoroughly explained.

\end{abstract}

\maketitle

\thispagestyle{fancy}


\section{Introduction}
Standard cosmology is based on the assumption that the Universe is spatially homogeneous, at least on scales sufficiently large to justify its approximation by a FLRW model. \\

However, the consensus on a 
homogeneous feature of structures, even on very large scales, 
has never been complete. At small separations, data worked out  using the correlation function method show a correlation lenght $r_0=5h^{-1}$ Mpc and the galaxy distribution exhibits a fractal structure with dimension $D_2\sim 1.2$~\cite{MD83}. The correlation integral method~\cite{PHC88} gives a fractal dimension slightly larger, $D_2 \sim 1.3$ to $1.5$. At larger scales, the value $D_2 \sim 2$ has been proposed~\cite{LG91} up to scales of at least $150h^{-1}$Mpc~\cite{FSL98}. For review articles see Martinez, 1999~\cite{VJM99} or Wu, Lahav and Rees, 1999~\cite{KW99}. It has also been suggested that three scaling regimes might  be discerned~\cite{GM98}. \\

At least, the value of the transition scale from inhomogeneity to homogeneity needs to be tested~\cite{MNC00,MNC05}. \\ 

We use the more recent and complete three-dimensional galaxy catalog, the Sloan Digital Sky Survey (SDSS), to repeat older calculations and hope to obtain more reliable results. \\

We  leave out of the account some other issues related to the galaxy case. These are source evolution and cosmological effects~\cite{RS80,MBR95,MNC01}. They should be considered in detail in some further study. \\

\section{The correlation Integral method}

As a first approach, we have chosen to use a characterization of the structures of point sets which is given by the correlation integral~\cite{PG83}, defined as:  
 \begin{equation}
C_2(r)= {1\over {N^{\prime}(N-1)}}\sum_i \sum_{j\ne i}
\Theta(r - |{\bf X}_i - {\bf X}_j|) 
\label{cor}
\end{equation}
where $\Theta$ is the Heaviside function. The inner summation is over the whole set of $N-1$ galaxies with coordinates ${\bf X}_j$, 
$j\ne i$, and the outer summation is over a subset of $N^{\prime}$ 
galaxies, taken as centers, with coordinates ${\bf X}_i$. By 
taking only the inner $N^{\prime}$ galaxies as centers we allow 
for the effect of the finiteness of the sample~\cite{FSL98}. \\

This characterization is also valid when the set is not fractal. Therefore it seems appropriate to use this approach to analyse  galaxies considered as point sets, provided the spaned scale range shows either a fractal behavior or not. \\

We may interpret $C_2(r)$ as ${\cal N}(r)/N$ where ${\cal N}(r)$ is the average number of galaxies within a distance $r$ of a typical galaxy in the set. As $r$ goes to zero, $C_2$ should vanish as $C_2 \propto r^{D_2}$. For computational purposes it is more convenient to use the form:
\begin{equation}
\log (C_2)=CONST. +D_2 \log (r) 
\label{lsq}
\end{equation}

The exponent $D_2$ is the fractal dimension, necessarily $\le 3$ for an embedding space of dimension three. When $D_2$ is different from three, the distribution is fractal~\cite{BM82}. 

\section{The analysis}

We use the publicly available SDSS data. After appropriate eliminations we are left with 84k galaxies. 
The rms galaxy redshift errors are estimated to be about 30km/sec~\cite{MT04}, therefore they are small for
the overall density fluctuations.  \\

We assume $H_0 = 65$ km/s/Mpc. Since our results exhibit large error bars, the value of $H_0$ is not too important. Another choice would only shift $h^{-1}$ the transition scales. \\

Because of the large differences between the transitions 
from one dimension to another, we devide the appropriate range  
into a large number of segments and calculate $D_2$ by the Procaccia-Grassberger method~\cite{PG83}. To obtain a very rough estimate on the region of transition, we check three points at a while. It gives the values for $D_2$ which are shown in Figures 1 and 2. \\

\begin{figure}
\includegraphics[width=65mm]{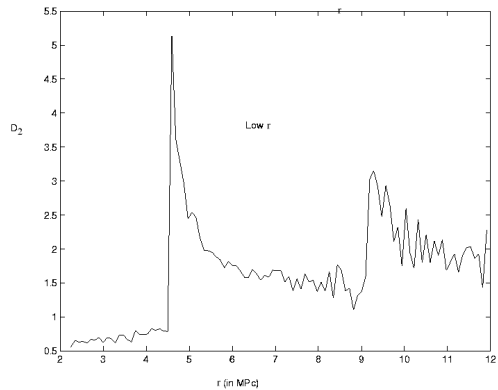}
\caption{$D_2$ as a function of $r$, for small scales.}
\label{Fig.1}
\end{figure}

\begin{figure}
\includegraphics[width=65mm]{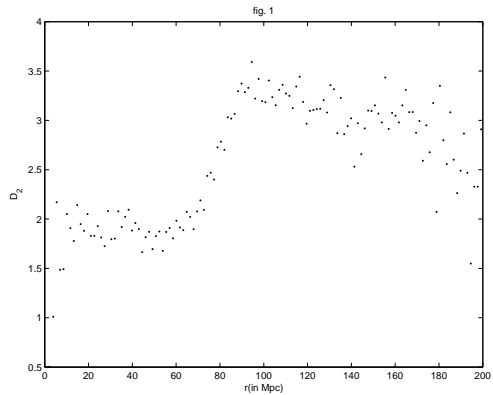}
\caption{$D_2$ as a function of $r$, for larger scales.}
\label{Fig.2}
\end{figure}

In the above figures, a step variation of the fractal dimension $D_2$ is obvious. This could explain why previous studies, limited in scale range, concluded to constant 
fractal dimensions with different values. However, large fluctuations appear at each transition scale. Then $D_2$  decreases towards a fluctuating value and rises again with large fluctuations at the next transition scale. The fluctuations are such that the fractal dimension becomes larger than the limiting dimension three at each transition. This seems to point out to some artefacts or bias due either to the employed method or to the data sample. The interpretations we propose are twofold.

\subsection{First interpretation (tentative)}

If one puts aside the low $r$ results of Figure \ref{Fig.1} because of the too unphysical values attained by $D_2$, one can use only the results of Figure \ref{Fig.2} and proceed to their following processing. \\

One devides the range from 0 to 200 Mpc into 256 segments. To obtain a rough estimate of the region of transition, one first checks three points at a while. This gives 128 values for $D_2$ (see Figure \ref{Fig.2}). The error on the average of the so obtained $D_2$ is taken as the error (more reliable than the (smaller) error obtained from the direct least square fit), because the distribution of these segment values is more Gaussian than the total least square fit. This method tends to give large errors on each segments but enables to observe more clearly the trend. Then one calculates the dimensions by using all the points to perform a least square calculation. \\

To check these results, we used two different
amounts of inner points in the evaluation. Still we kept in mind
that all the points, including those outer points which are at
a distance used in our calculations, should be within the measured 
range. We also used, as the maximal distance considered, three 
different values. In all the cases the transition area remained 
the same. We then took an inner region which is much larger so
that some of the points of the outer region are already in an area 
not covered by the catalogue. For this case the results differed 
considerably. We do not get a clear transition and the dimension
was on the average $2.3 \pm 0.3$. The trend is obvious as less and
less points are counted from the total available points set
we expect to get more and more distorted results. \\

\begin{figure}
\includegraphics[width=65mm]{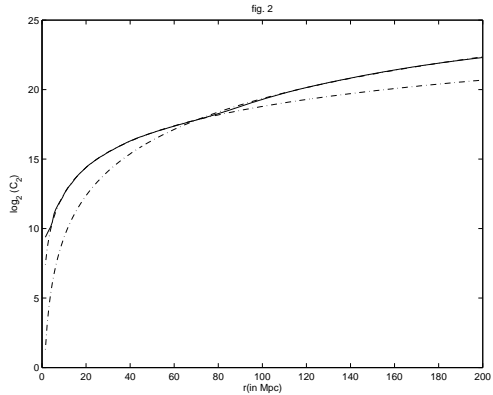}
\caption{The tentative $D_2=2$ to $D_2=3$ transition. Taking the parameters of the least square fit for the lower part and upper part, compared to the experimental results (solid line).}
\label{Fig.3}
\end{figure}

\begin{figure}
\includegraphics[width=65mm]{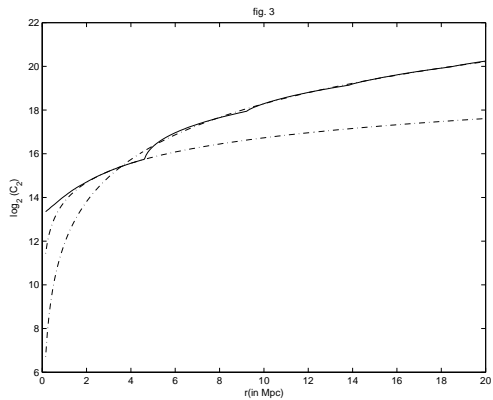}
\caption{The tentative $D_2\approx 1$ to $D_2=2$ transition with the same procedure as for Figure 3.}
\label{Fig.4}
\end{figure}

In Figure \ref{Fig.3}, a transition seems to appear between two scaling regimes: \\
$1.90 \pm 0.03$ , for 6 Mpc to 80 Mpc. \\
$3.01 \pm 0.04$ , for 100 Mpc to 200 Mpc. \\

Between 80 Mpc and 100 Mpc the picture is not clear, so that the transition might be somewhere around 90 Mpc. In Figure \ref{Fig.4}, which is obtained by the same method applied to the low $r$ results of Figure \ref{Fig.1}, another transition seems to appear 
between a $D_2 \sim 1$ and $D_2 \sim 2$ regime: \\
$0.89 \pm 0.01$ for 2 Mpc to 4.5 Mpc. \\
$1.93 \pm 0.05$ for 6 Mpc to 20 Mpc. \\

Between 4.5 Mpc and 6 Mpc the picture is also not too clear. At the transition, the dimension increases very
dramatically. This might be interpreted as an artefact resulting from the small difference in $r$ of neighbouring bins. 
But these low $r$ analyses can also be viewed as some confirmation of the preliminary results obtained by~\cite{GM98} who found, for galaxy separations up to about 5 Mpc a $D_2$ dimension about 1.2 and for larger scales (out to about 30 Mpc, which was the limiting scale of their study) a dimension of about 1.8.

\subsection{Second interpretation (seemingly more robust, but to be confirmed)}

Figures 1 and 2 show a similar behavior, 
with a peak at the transition, then an ($e^{-r}$)-like decrease toward a constant value (with fluctuations above the allowed value $D_2=3$). When more data become available, and with an analysis method more adapted to the study of multi-fractal distributions, we might be able to check whether this peak is actually an artefact coming from our data analysis method (some step behavior at each transition scale) or if it is due to the data sample. \\

The $D_2=3.3$ peak at 100 Mpc in \ref{Fig.2} and the decrease farther could be of the same nature: in this case, the ``true'' value of $D_2$  beyond 100 Mpc would not yet have been reached at the limit of the study (200 Mpc) (it might be $\leq 2.6$), which could mean that the transition to uniformity ($D_2=3$) has not yet been reached at these scales.

\section{Conclusion and discussion}

We use the publicly available data from the SDSS, to 
complete an analysis of the fractal dimension of the galaxy 
distribution, with the correlation integral method. \\

We check scales up to 130 $h^{-1}$ Mpc. We obtain an obvious step variation of the fractal dimension $D_2$. This could explain why previous studies, limited in scale range, concluded to constant fractal dimensions, with different values. \\

Two possible interpretations of these results are proposed:
\begin{itemize}
 \item A rough mean square fit method 
gives i) up to 4.5 to 6 Mpc, a dimension of the order one, 
ii) then a transition to a dimension of the order two, 
iii) and between 80 and 100 Mpc, another transition to a 
dimension around three.
 \item However, the variation of $D_2$ with scales show a peak 
 at each transition (the fractal dimensions at the peaks become 
 larger than 3), then an ($e^{-r}$)-like decrease toward a constant value (with fluctuations). This could be due to the step behavior at the transition scale. The transition to uniformity ($D_2=3$) would thus not yet have been reached at the largest studied scales. Moreover, a transition to homogeneity at 130  $h^{-1}$ Mpc would be inconsistent with the sizes of the
largest structures seen in the universe ~\cite{WS91,JE03}. Last, it would be interesting to check 
if a variable with scale fractal dimension of the galaxy 
distribution might be related to theoretical 
predictions proposed as a consequence of a principle of 
relativity of scales~\cite{LN95}: a transition 
to homogeneity predicted around 750 Mpc and a multi-fractal distribution with a dimension varying by steps whose width might be related to the organization hierarchy observed for the galaxies.\\
\end{itemize}

 However, these large fluctuations, which appear at each transition scale, seem to point out to some artefacts due  either to the employed method or to the data sample. A way to discriminate between the two possible reasons of the appearance of these odd fluctuations would be to test the validity of the application of the Grassberger and Procaccia's correlation integral method to a mutifractal distribution by applying it to a set of mock catalogues of galaxies, artificially constructed, with a known fractal dimension varying by known steps. If the runs of the here employed code reproduce the known features of the distributions, it would suggest that the artefacts might be due to the data sample analysed here. We would therefore need a better sample to go further on and, e. g., apply our method to the next SDSS catalogue, presumably publicly available in a very near future. If, on the contrary, those runs are not able to reproduce the known features of the distributions, it would suggest that the artefacts might be due to the employed method, and we would have to test other ones and find a code better fitted to the study of multifractal distributions.\\

{\it Aknowledgements.}  
The authors wish to thank Professors E.A. Spiegel and L. Nottale for valuable suggestions and discussions.

\end{document}